\documentclass[12pt]{article}
\usepackage{amssymb,amsmath,amsbsy}
\usepackage{bbm,latexsym,epsfig}

\newcommand{\be}{\begin{equation}}
\newcommand{\ee}{\end{equation}}
\newcommand{\ba}{\begin{eqnarray}}
\newcommand{\ea}{\end{eqnarray}}
\newcommand{\no}{\nonumber\\}

\newcommand{\deltasol}{\Delta m^2_\mathrm{sol}}
\newcommand{\deltaatm}{\Delta m^2_\mathrm{atm}}

\begin{document}

\title{\normalsize \hfill CFTP/12-003 \\[1cm]
\LARGE
Texture-zero model for the lepton mass matrices}

\author{
P.~M.~Ferreira$^{(1,2)}$\thanks{E-mail: ferreira@cii.fc.ul.pt} \ and \
L.~Lavoura$^{(3)}$\thanks{E-mail: balio@cftp.ist.utl.pt} \\
\\*[2mm]
\small $^{(1)}$ Instituto Superior de Engenharia de Lisboa,
1959-007 Lisboa, Portugal
\\*[2mm]
\small $^{(2)}$ Centre for Theoretical and Computational Physics,
University of Lisbon \\
\small 1649-003 Lisboa, Portugal
\\*[2mm]
\small $^{(3)}$ Technical University of Lisbon and CFTP \\
\small Instituto Superior T\'ecnico, 1049-001 Lisboa, Portugal
\\*[5mm]
}

\date{23 July 2012}

\maketitle

\begin{abstract}
We suggest a simple model,
based on the type-I seesaw mechanism,
for the lepton mass matrices.
The model hinges on an Abelian symmetry
which leads to mass matrices with some vanishing matrix elements.
The model predicts one massless neutrino and $M_{e\mu} = 0$
($M$ is the effective light-neutrino Majorana mass matrix).
We show that these predictions
agree with the present experimental data
if the neutrino mass spectrum is inverted,
\textit{i.e.}\ if $m_3 = 0$,
provided the Dirac phase $\delta$ is very close to maximal
($\pm \pi / 2$).
In the case of a normal neutrino mass spectrum,
\textit{i.e.}\ when $m_1 = 0$,
the agreement of our model with the data is
imperfect---the
reactor mixing angle $\theta_{13}$ is too small in our model.
Minimal leptogenesis is not an option in our model
due to the vanishing elements in the Yukawa-coupling matrices.

\vspace*{10mm}

\end{abstract}

\newpage

\paragraph{Notation:}
In the model of this Letter we utilize the type-I seesaw mechanism.
There are right-handed neutrinos $\nu_R$
which are invariant under the $SU(2) \times U(1)$ gauge group.
One defines the left-handed conjugates $\nu^\prime_L$ of the $\nu_R$ through
\be
\nu^\prime_L = C \bar \nu_R^T,
\ee
where $C$ is the Dirac--Pauli charge-conjugation matrix.
The neutrino mass terms are given by
\begin{subequations}
\ba
\mathcal{L}_{\nu \, \mathrm{mass}} &=&
- \bar \nu_R M_D \nu_L
- \bar \nu_L M_D^\dagger \bar \nu_R
- \frac{1}{2}\, \bar \nu_R M_R C \bar \nu_R^T
+ \frac{1}{2}\, \nu_R^T C^{-1} M_R^\ast \nu_R
\\ &=&
\frac{1}{2}\, \left( \begin{array}{cc}
\nu_L^T, & {\nu_L^\prime}^T \end{array} \right)
C^{-1} \left( \begin{array}{cc}
0 & M_D^T \\ M_D & M_R \end{array} \right)
\left( \begin{array}{c}
\nu_L \\ \nu_L^\prime \end{array} \right)
+ \mathrm{H.c.},
\label{wbuti}
\ea
\end{subequations}
where,
without loss of generality,
the matrix $M_R$ is symmetric.
Under the conditions of the seesaw mechanism,
the approximate Majorana mass matrix for the light neutrinos is then
\be
M = - M_D^T M_R^{-1} M_D.
\label{svuit}
\ee
The symmetric matrix $M$ is bi-diagonalized by a unitary matrix $U$ via
\be
U^T M U =
\mathrm{diag} \left( \mu_1, \ \mu_2, \ \mu_3 \right),
\ee
where $\mu_j = m_j e^{2 i \beta_j}$
($j = 1, 2, 3$, no sum over $j$);
the (real and non-negative) $m_j$ are the neutrino masses
and the $\beta_j$ are their respective Majorana phases.
The matrix $U$ may be parametrized
\be
U =
\mathrm{diag} \left(
e^{i \gamma_e}, \ e^{i \gamma_\mu}, \ e^{i \gamma_\tau} \right)
\left( \begin{array}{ccc}
c_{12} c_{13} & s_{12} c_{13} & s_{13} e^{- i \delta} \\
- s_{12} c_{23} - c_{12} s_{23} s_{13} e^{i \delta} &
c_{12} c_{23} - s_{12} s_{23} s_{13} e^{i \delta} &
s_{23} c_{13} \\
s_{12} s_{23} - c_{12} c_{23} s_{13} e^{i \delta} &
- c_{12} s_{23} - s_{12} c_{23} s_{13} e^{i \delta} &
c_{23} c_{13}
\end{array} \right),
\ee
where the $\gamma_\alpha$ ($\alpha = e, \mu, \tau$) are unphysical phases,
$c_i = \cos{\theta_i}$
($i = 12, 23, 13$),
$s_i = \sin{\theta_i}$,
and $\delta$ is the physical (observable) `Dirac' phase.
The angles $\theta_i$ belong to the first quadrant.

\paragraph{Purpose:}
Recently~\cite{Lavoura:2011ry},
one of us has suggested a model that yields,
in a certain limit and in the basis where
the charged-lepton mass matrix $M_\ell$ is diagonal,
the predictions
\begin{subequations}
\label{uviwe}
\ba
\det{M} &=& 0, \label{deter}
\\
M_{e\mu} &=& 0. \label{m12}
\ea
\end{subequations}
It has been claimed~\cite{Lavoura:2011ry}
that the predictions~(\ref{uviwe})
``are compatible with the experimental data
on neutrino masses and on lepton mixing''
but this assertion has not been proved and qualified
in ref.~\cite{Lavoura:2011ry}.
It is the aim of this Letter to
\begin{enumerate}
\item Present a model,
simpler than the one in ref.~\cite{Lavoura:2011ry},
that leads to the predictions~(\ref{uviwe}).
\item Work out analytically
the consequences of those predictions
for the parameters of the matrix $U$,
both in the cases of an inverted and of a normal neutrino mass spectrum.
\item Display graphically those predictions by means of scatter plots.
\item Analyse the consequences of our model for leptogenesis.
\item Show that,
in the context of our model,
the conditions~(\ref{uviwe}) are stable
under the (one-loop) renormalization-group evolution of $M$.
\end{enumerate}
To be sure,
the conditions~(\ref{uviwe}) have been studied in the literature before,
see for instance refs.~\cite{Ibarra:2003up,Lashin:2011dn}.
However,
the points 1,
4,
and 5 above are original,
and we believe that we have also performed
the tasks 2 and 3 in a simpler,
more transparent fashion than in the previous literature.

\paragraph{Mass matrices:}
We shall assume that
\emph{there are only two right-handed neutrinos}
$\nu_R$~\cite{Ibarra:2003up}.
This means that the matrix $M_D$ is $2 \times 3$ and,
therefore,
the $5 \times 5$ mass matrix in equation~(\ref{wbuti})
has a $3 \times 3$ null submatrix,
hence it has (at least) one zero eigenvalue,
\textit{i.e.}\ \emph{there is one massless neutrino}.
Furthermore,
in the basis where $M_R$ is diagonal,
\be
M_R =
\mathrm{diag} \left( x, \ y \right),
\label{mr}
\ee
\textit{the matrix $M_D$ has two texture zeros}:
\be
M_D = \left( \begin{array}{ccc}
a & 0 & c \\ 0 & b & d
\end{array} \right).
\label{md}
\ee
By utilizing equation~(\ref{svuit}),
one then has
\be
M = - \left( \begin{array}{ccc}
a^2 / x & 0 & a c / x \\ 0 & b^2 / y & b d / y \\ a c / x & b d / y &
c^2 / x + d^2 / y \end{array} \right).
\ee
This matrix features equations~(\ref{uviwe}).
Notice that we are implicitly assuming that
the charged-lepton mass matrix is diagonal too.
The condition~(\ref{m12}) means that
\be
\mu_1^\ast U_{11} U_{21} + \mu_2^\ast U_{12} U_{22} + \mu_3^\ast U_{13} U_{23} = 0.
\label{jvuti}
\ee
The condition~(\ref{deter}) means that one of the neutrinos is massless.
Thus,
either $\mu_ 1 = 0$ and then the neutrino mass spectrum is normal,
with $m_2 = \sqrt{\deltasol}$ and $m_3 = \sqrt{\deltaatm}$,
or $m_3 = 0$ and then the neutrino mass spectrum is inverted,
with $m_1 = \sqrt{\deltaatm}$ and $m_2 = \sqrt{\deltaatm + \deltasol}$.

\paragraph{Inverted spectrum:}
In the inverted-spectrum case,
equation~(\ref{jvuti}) implies\footnote{The relative phases
of the terms in the left-hand side of equation~(\ref{jvuti})
depend on the Majorana phases and are therefore of little relevance.}
\ba
& & m_1^2 c_{12}^2 \left( s_{12}^2 c_{23}^2
+ c_{12}^2 s_{23}^2 s_{13}^2
+ 2 c_{12} s_{12} c_{23} s_{23} s_{13} \cos{\delta} \right)
\no &=&
m_2^2 s_{12}^2 \left( c_{12}^2 c_{23}^2
+ s_{12}^2 s_{23}^2 s_{13}^2
- 2 c_{12} s_{12} c_{23} s_{23} s_{13} \cos{\delta} \right).
\ea
Therefore,
\ba
0 &=& \left( m_2^2 - m_1^2 \right) c_{12}^2 s_{12}^2 c_{23}^2
+ \left( m_2^2 s_{12}^4 - m_1^2 c_{12}^4 \right) s_{23}^2 s_{13}^2
\no & &
- \left( m_2^2 s_{12}^2 + m_1^2 c_{12}^2 \right)
\left( 2 c_{12} s_{12} c_{23} s_{23} s_{13} \cos{\delta} \right).
\label{uvity}
\ea
One may compute the approximate value of each term
in the right-hand side of equation~(\ref{uvity}) by using as a guide
the \textit{Ansatz}\/ of tri-bimaximal mixing~\cite{Harrison:2002er}
$s_{12} = 3^{-1/2}$,
$s_{23} = 2^{-1/2}$.
Thus, equation~(\ref{uvity}) may be approximated by
\be
0 \approx \frac{\deltasol}{9} - \frac{\deltaatm}{6}\, s_{13}^2
- \frac{\sqrt{2} \deltaatm}{3}\, s_{13} \cos{\delta}.
\label{wvuit}
\ee
We use in this Letter the phenomenological values given by~\cite{schwetz}.
A different phenomenological fit to the experimental data is the one 
by~\cite{sch2}.
Numerically,
using $s_{13}^2 \approx 0.024$,
the first term in the right-hand side of equation~(\ref{wvuit}) is about
$8.4 \times 10^{-6}$ eV$^2$,
the second term is about
$- 9.72 \times 10^{-6}$ eV$^2$,
and the coefficient of $\cos{\delta}$ in the third term is about
$1.77 \times 10^{-4}$ eV$^2$.
One concludes from this estimate that equation~(\ref{wvuit}) implies
\be
\cos{\delta} \approx 0.
\label{cosseno}
\ee
Remarkably, equation~(\ref{uvity})
does not imply any constraint on the mixing angles---it works perfectly well
for any of their phenomenologically allowed values.
One concludes that,
in the case of an inverted neutrino mass spectrum,
our conditions~(\ref{uviwe})
fit very well\footnote{A a matter of fact,
our model leads to $\delta \sim \pi / 2$, while
ref.~\cite{schwetz} prefers $\cos{\delta} \approx -1$,
but only at the $1 \sigma$ level;
at $2 \sigma$, ref.~\cite{schwetz} leaves
$\delta$ free.}
the phenomenological values of the mixing angles
and make the prediction~(\ref{cosseno}) for the Dirac phase.

\paragraph{Normal spectrum:}
In the normal-spectrum case,
equation~(\ref{jvuti}) implies
\be
\deltasol s_{12}^2 \left( c_{12}^2 c_{23}^2
+ s_{12}^2 s_{23}^2 s_{13}^2
- 2 c_{12} s_{12} c_{23} s_{23} s_{13} \cos{\delta} \right)
= \deltaatm s_{23}^2 s_{13}^2.
\label{liyt}
\ee
In the left-hand side of this equation,
the first term inside the parenthesis dominates over the other two
because $s_{23} \approx c_{23}$,
$c_{12} > s_{12}$,
and $s_{13} \ll 1$.
Therefore,
\be
\frac{\deltasol}{\deltaatm} \approx \frac{s_{23}^2}{c_{23}^2}\,
\frac{s_{13}^2}{c_{12}^2 s_{12}^2}.
\label{uvyti}
\ee
Phenomenologically,
the left-hand side of equation~(\ref{uvyti}) is about $0.03$,
while in the right-hand side
$s_{23}^2 \left/ c_{23}^2 \right. \approx 1$
and $1 \left/ \left( c_{12}^2 s_{12}^2 \right) \right. \approx 4.5$.
Therefore,
the model predicts $s_{13}^2 \approx 0.0067$,
which is
much
below the best-fit value for $s_{13}^2$~\cite{schwetz}.
Moreover,
while $1 \left/ \left( c_{12}^2 s_{12}^2 \right) \right.$
is rather stable over the $\theta_{12}$ allowed range,
$s_{23}^2 \left/ c_{23}^2 \right.$ lies at $3\, \sigma$
in between 0.49 and 1.75,
which means that
$s_{13}^2$ must be significantly correlated with $s_{23}^2$.
One concludes that,
if the neutrino mass spectrum is normal,
then the condition~(\ref{m12}) does not give a perfect fit to the data
because it implies a much too low $\theta_{13}$;
moreover,
there should be a correlation between the mixing angles $\theta_{13}$
and $\theta_{23}$.
The phase $\delta$ remains,
in the case of a normal neutrino mass spectrum,
free.

\paragraph{Neutrinoless $2 \beta$ decay:}
One is particularly interested in the quantity
$m_{\beta \beta} \equiv \left| M_{ee} \right|$,
which is the mass relevant for neutrinoless double-beta decay.
One easily finds that equation~(\ref{jvuti})
together with the unitarity of $U$ imply
\ba
m_1 = 0 &\Rightarrow& m_{\beta \beta} =
m_3 \left| \frac{U_{13} U_{31}}{U_{22}} \right|
\sim m_3 \frac{s_{13}}{\sqrt{2}},
\label{m1} \\
m_3 = 0 &\Rightarrow& m_{\beta \beta} =
m_1 \left| \frac{U_{11} U_{33}}{U_{22}} \right|
\sim m_1,
\label{m3}
\ea
where in the second step we have taken into account the approximate values
of the matrix elements of $U$
given by the \textit{Ansatz}\/ of tri-bimaximal mixing.
One sees that $m_{\beta \beta} \sim \sqrt{\deltaatm} \approx 0.05$~eV
in the inverted-spectrum case,
and that
$m_{\beta \beta} \sim \left. \sqrt{\deltaatm} \right/ \! 10 \approx 0.005$~eV
in the normal case~\cite{Branco:2002ie}~\footnote{These are,
as a matter of fact,
typical values for $m_{\beta\beta}$
which follow solely from the condition~(\ref{deter})
and independently of the condition~(\ref{m12}),
as was shown in~\cite{Branco:2002ie}.}.
Thus,
in our model $m_{\beta \beta}$ might be measurable
if the neutrino mass spectrum is inverted,
but is certainly too low to be measured
(at least any time soon) if the mass spectrum is normal.

\begin{figure}[ht]
\begin{center}
\mbox{\epsfig{file=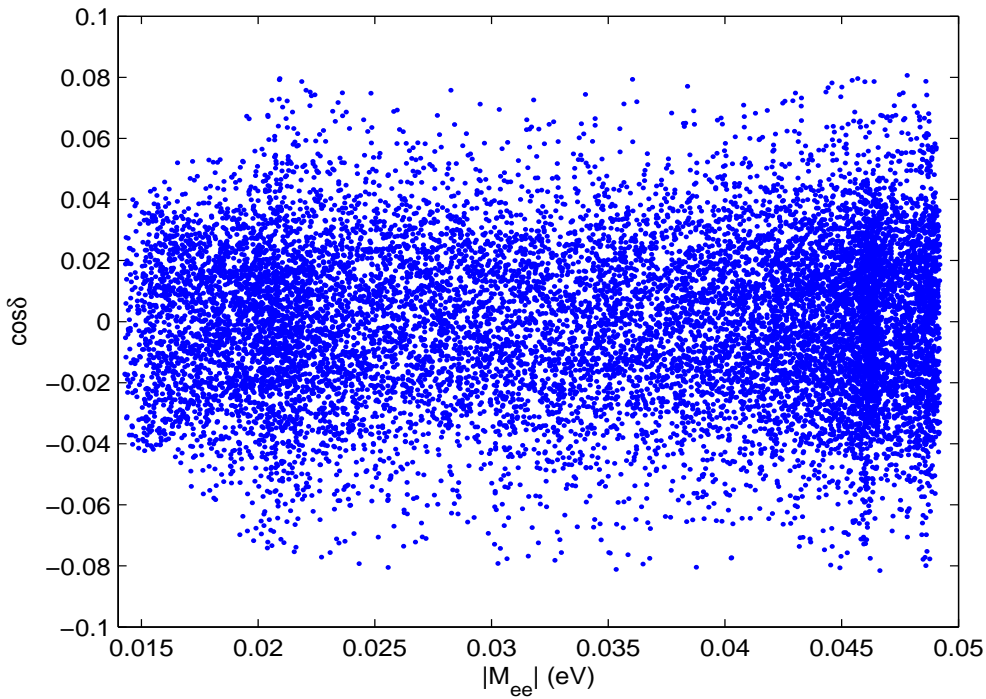,
width=0.9\textwidth,
height=10cm}}
\end{center}
\caption{Scatter plot of the Dirac phase $\delta$ against
$m_{\beta\beta} = |M_{ee}|$
in the case of an inverted hierarchy of neutrino masses.
The fit is made to the
2$\sigma$
data of ref.~\cite{schwetz}.
Notice that the fit strongly pushes $\delta$ to its maximal values.}
\label{fig:inv}
\end{figure}

\paragraph{Numerical analysis:}
We now undertake a scan of the parameter space
allowed by the most recent phenomenological data
on neutrino masses and oscillations~\cite{schwetz}.
As explained earlier,
in the inverted-hierarchy case
our model fits the data quite well,
so we only had to use for the observables
the 2$\sigma$ intervals in ref.~\cite{schwetz}.
The fitting procedure consisted in varying the several observables
(the angles $\theta_{12}$,
$\theta_{13}$,
and $\theta_{23}$,
the mass parameters $m_1$ and $m_2$,
and the phase $\delta$)
over their allowed ranges,
but making them obey equation~\eqref{uvity}.
Remarkably,
this fit to the experimental data
enforces no constraints on the parameters of the matrix $U$,
with the exception of the phase $\delta$,
which is found to be close to its maximal value $\pm \pi/2$,
as seen in figure~\ref{fig:inv}.
In that figure,
we plot the quantity $m_{\beta\beta} = |M_{ee}|$,
relevant for neutrinoless double-beta decay,
versus the cosine of $\delta$.
Each point in that plot
corresponds to a particular combination of $U$-matrix parameters
which obey the constraint of equation~\eqref{uvity}.
We thus confirm the two estimates shown previously:
that the CP-violating phase $\delta$ is close to
$\pm \pi / 2$,
and that $m_{\beta\beta} \sim$ 0.05 eV.

\begin{figure}[ht]
\centerline{\epsfysize=6cm
\epsfbox{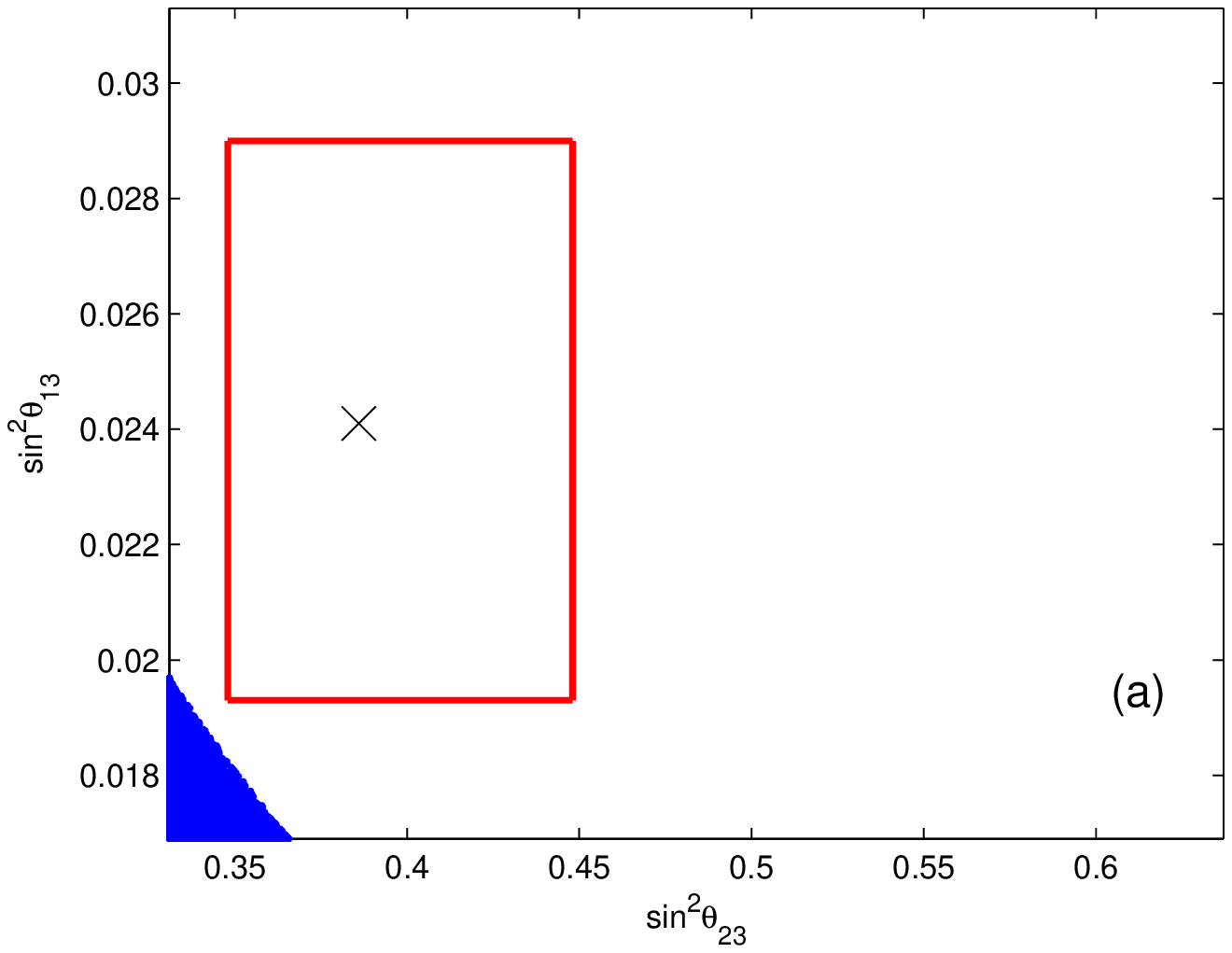}\epsfysize=6cm \epsfbox{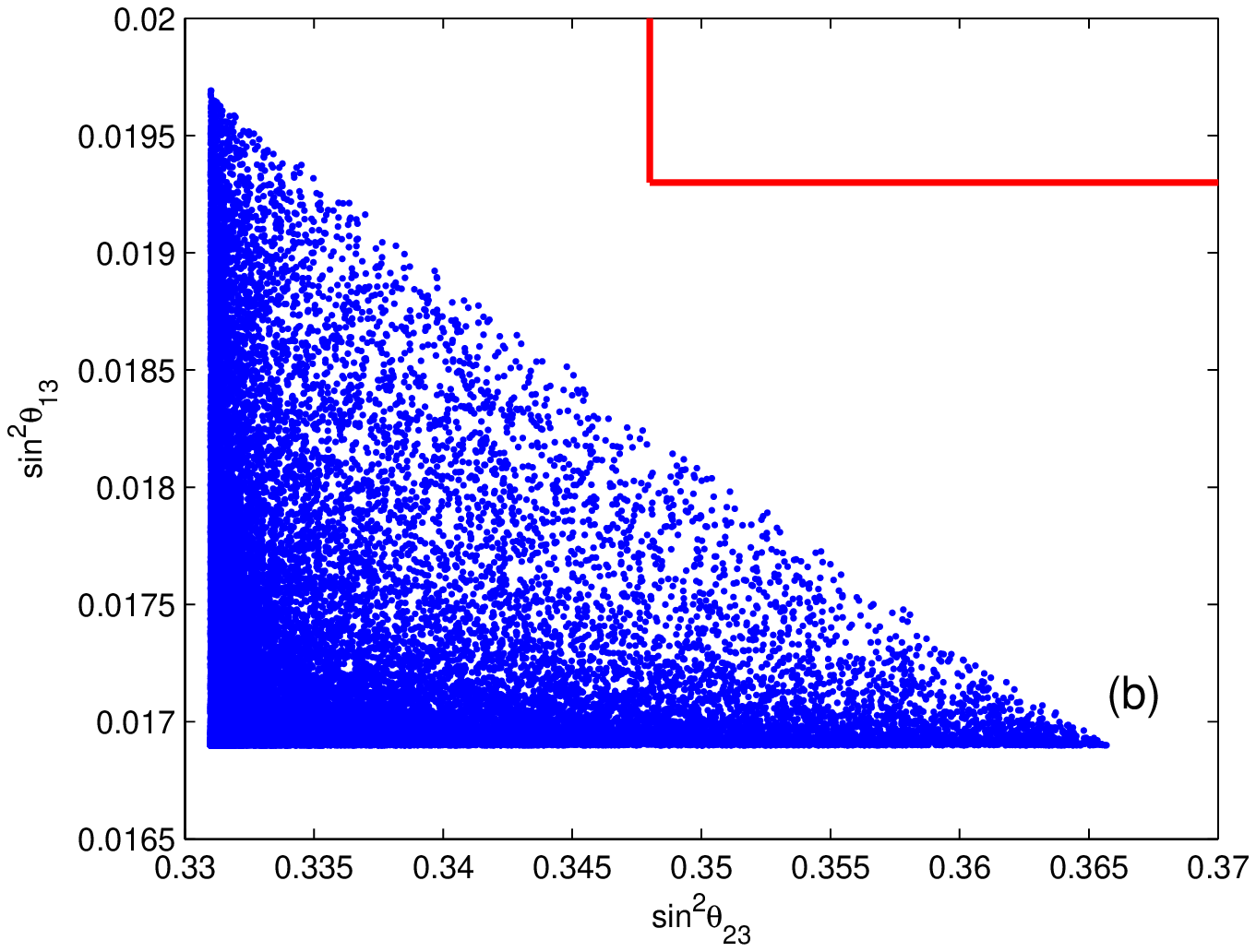} }
\caption{Scatter plot of the reactor angle $\theta_{13}$
against the atmospheric angle $\theta_{23}$,
for the case of normal hierarchy of the neutrino masses.
The fit is made to the 3$\sigma$ data of ref.~\cite{schwetz}.
The boundaries of figure 2(a) correspond to the $3 \sigma$ bounds,
the red lines mark the $2 \sigma$ bounds,
and the cross indicates the best-fit point.
Figure 2(b) is the magnification of the low-left corner of figure 2(a).
Notice that the fit pushes both $\theta_{13}$ and $\theta_{23}$ to small values,
in such a way that at $2 \sigma$ level the model is excluded.}
\label{fig:norm}
\end{figure}

The fitting procedure is analogous for the normal-hierarchy case,
except that we now want the parameters to obey equation~\eqref{liyt}
instead of equation~(\ref{uvity}).
It turns out that
\emph{the fit is impossible if one uses the 2$\sigma$ intervals}\
for the various
observables;
we have therefore used the 3$\sigma$ intervals instead.
The main problem is with the angle $\theta_{13}$,
which is driven by the condition~\eqref{liyt} to values much too small.
This can be appreciated in figure~\ref{fig:norm},
where we have plotted $\sin^2{\theta_{13}}$ against $\sin^2{\theta_{23}}$.
The red cross indicates the phenomenological best-fit value;
one sees that our model yields $\theta_{13}$ always well below
that best-fit value.
In figure~\ref{fig:norm} one can also observe
the correlation between the $\theta_{13}$ and $\theta_{23}$
mentioned earlier.
We have found no other correlations between observables
in this normal-hierarchy case.
Unlike in the inverted-hierarchy case,
the value of $m_{\beta\beta}$ is found to be always extremely small---in between
0.0017 eV and 0.0041 eV---and
the CP-violating phase $\delta$ can take any value.

\paragraph{A model:}
We now explicitly construct a model which generates the neutrino mass matrices
in equations~(\ref{mr}) and~(\ref{md})
together with a diagonal charged-lepton mass matrix.
Those mass matrices are solely characterized
by some zero matrix elements,
and it is known~\cite{Grimus:2004hf} that
any set of texture zeros can always be justified through a model
furnished with an \emph{Abelian}\ flavour symmetry and,
possibly,
a large number of scalar gauge multiplets.
Indeed,
there are always many such possible models;
the model that we shall present in this paragraph
is only one (possibly minimal in terms of its scalar content)
out of many others which might perform the job.
Our model is based
on the Abelian symmetry $\mathbbm{Z}_8$;
let $\sigma = \sqrt[8]{1} = \exp{\left( i \pi / 4 \right)}$.
The model has three scalar gauge-$SU(2)$ doublets
$\Phi_b = \left( \phi_b^+, \, \phi_b^0 \right)^T$
($b = 1, 2, 3$)
and one complex gauge-invariant scalar $S$.
Under $\mathbbm{Z}_8$,
$\Phi_1$ is invariant and
\be
\Phi_2 \to \sigma^3 \Phi_2, \
\Phi_3 \to \sigma^6 \Phi_3, \
S \to \sigma^2 S, \
\nu_{R1} \to \sigma \nu_{R1}, \
\nu_{R2} \to \sigma^3 \nu_{R2}.
\ee
Therefore,
the scalar potential will contain the phase-sensitive terms
$\Phi_1^\dagger \Phi_2\, \Phi_3^\dagger \Phi_2$,
$\Phi_1^\dagger \Phi_3\, S$,
and $S^4$,
together with their Hermitian conjugates,
thereby avoiding the appearance of any Goldstone bosons
when $S$ and the $\phi_b^0$ acquire vacuum expectation values (VEVs).
The matrix $M_R$ is generated by the Yukawa couplings
of the right-handed neutrinos to $S$,
\be
\mathcal{L}_{S \, \mathrm{Yuk}} =
- \frac{y_1}{2}\, \bar \nu_{R1} C \bar \nu_{R1}^T S
- \frac{y_2}{2}\, \bar \nu_{R2} C \bar \nu_{R2}^T S^\ast
+ \mathrm{H.c.},
\ee
yielding $x = y_1 s$ and $y = y_2 s^\ast$,
where $s = \left\langle 0 \left| S \right| 0 \right\rangle$ is the VEV of $S$.
The left-handed lepton doublets
$D_{L\alpha} = \left( \nu_{L\alpha}, \, \alpha_L \right)^T$
and the right-handed charged-lepton singlets $\alpha_R$
transform under $\mathbbm{Z}_8$ as
\be
D_{Le} \to \sigma D_{Le}, \
D_{L\mu} \to \sigma^5 D_{L\mu}, \
D_{L\tau} \to \sigma^3 D_{L\tau}, \
e_R \to \sigma e_R, \
\mu_R \to \sigma^7 \mu_R, \
\tau_R \to \tau_R.
\ee
This leads to the following $\Phi_b$ Yukawa Lagrangian:
\ba
\mathcal{L}_{\Phi \, \mathrm{Yuk}} &=&
- \left( \phi_1^0, \, - \phi_1^+ \right)
\left( y_3 \bar \nu_{R1} D_{Le} + y_4 \bar \nu_{R2} D_{L\tau} \right)
- \left( \phi_3^0, \, - \phi_3^+ \right)
\left( y_5 \bar \nu_{R1} D_{L\tau} + y_6 \bar \nu_{R2} D_{L\mu} \right)
\no & &
- y_7 \left( \phi_1^-, \, {\phi_1^0}^\ast \right) \bar e_R D_{Le}
- y_8 \left( \phi_3^-, \, {\phi_3^0}^\ast \right) \bar \mu_R D_{L\mu}
- y_9 \left( \phi_2^-, \, {\phi_2^0}^\ast \right) \bar \tau_R D_{L\tau}
+ \mathrm{H.c.}
\label{jzirt}
\ea
Let $v_b = \left\langle 0 \left| \phi_b^0 \right| 0 \right\rangle$
denote the VEV of $\phi_b^0$,
then the charged-lepton masses are
$m_e = \left| y_7 v_1 \right|$,
$m_\mu = \left| y_8 v_3 \right|$,
and $m_\tau = \left| y_9 v_2 \right|$;
the elements of the neutrino Dirac mass matrix $M_D$ in equation~(\ref{md})
are $a = y_3 v_1$,
$b = y_6 v_3$,
$c = y_5 v_3$,
and $d = y_4 v_1$.
Notice that,
as mentioned earlier,
the charged-lepton mass matrix is automatically diagonal with this symmetry.
Furthermore,
unlike what usually happens in flavour models
based on non-Abelian
symmetries,
there is here no specific requirement
(beyond conservation of electromagnetism)
on the model's vacuum and,
therefore,
the fit to the leptonic sector imposes no constraints on the scalar sector.

\paragraph{Leptogenesis:}
We next consider
leptogenesis. (For reviews of leptogenesis see,
for instance, refs.~\cite{Paschos:2003jc,Branco:2011zb}.)
We firstly do this in the context of
a general multi-Higgs-doublet model (MHDM).
Let there be $n_H$ Higgs doublets $\Phi_b$
($b = 1, 2, \ldots, n_H$)
and $n_\nu$ right-handed neutrinos $\nu_{Ri}$
($i = 1, 2, \ldots, n_\nu$);
the latter are supposed to be eigenstates
of mass and, in this paragraph, $m_i$ denotes the mass of $\nu_{Ri}$.
The Yukawa couplings of the right-handed neutrinos
are given by
\be
\mathcal{L}_{\nu \, \mathrm{Yuk}} =
- \sum_{b=1}^{n_H} \sum_{i=1}^{n_\nu} \sum_{\alpha = e, \mu, \tau}
\left( Y_b \right)_{i\alpha} \bar \nu_{Ri}
\left( \phi_b^0, \, - \phi_b^+ \right)
\left( \begin{array}{c} \nu_{L\alpha} \\ \alpha_L^- \end{array} \right)
+ \mathrm{H.c.},
\ee
where the $Y_b$ are dimensionless $n_\nu \times 3$ Yukawa-coupling matrices.
In the approximation where the right-handed neutrinos are (very) massive
but both the charged leptons $\alpha^\pm$
and the charged scalars $\phi_b^\pm$ are massless,
the CP-violating asymmetry between the decay rates of $\nu_{Ri}$
to $\alpha^+ \phi_b^-$ and $\alpha^- \phi_b^+$,
summed over the flavour $\alpha$,
is given by
\begin{subequations}
\label{asym}
\ba
\epsilon_{ib} &=& \frac{1}{8 \pi \left( Y_b Y_b^\dagger \right)_{ii}}
\sum_{c=1}^{n_H}
\sum_{j=1}^{n_\nu}
\left\{
\mathrm{Im} \left[
\left( Y_c Y_b^\dagger \right)_{ji} \left( Y_b Y_c^\dagger \right)_{ji}
\right] f \left( x_j \right)
\right. \label{line1} \\ & & \left.
+ \mathrm{Im} \left[
\left( Y_b Y_b^\dagger \right)_{ji} \left( Y_c Y_c^\dagger \right)_{ji}
\right] g \left( x_j \right)
\right\},
\label{line2}
\ea
\end{subequations}
where
$x_j \equiv m_j^2 \left/ m_i^2 \right.$
and~\cite{Paschos:2003jc}
\begin{subequations}
\ba
f \left( x \right) &=& \sqrt{x} \left[ 1 + \left( 1 + x \right)
\ln{\frac{x}{1 + x}} \right],
\\
g \left( x \right) &=& \frac{\sqrt{x}}{1 - x}.
\ea
\end{subequations}
We now consider the particular case of our model in the previous paragraph.
There,
\be
Y_1 =
\left( \begin{array}{ccc}
y_3 & 0 & 0 \\ 0 & 0 & y_4
\end{array} \right),
\quad
Y_2 = \left( \begin{array}{ccc}
0 & 0 & 0 \\ 0 & 0 & 0
\end{array} \right),
\quad
Y_3 =
\left( \begin{array}{ccc}
0 & 0 & y_5 \\ 0 & y_6 & 0
\end{array} \right).
\label{wvuir}
\ee
It is easy to see that
the matrices $Y_1 Y_1^\dagger$ and $Y_3 Y_3^\dagger$ are diagonal,
and therefore do not contribute to line~(\ref{line2}),
while $\left( Y_1 Y_3^\dagger \right)_{12} = 0$
and $\left( Y_3 Y_1^\dagger \right)_{21} = 0$,
and therefore line~(\ref{line1}) vanishes too.
We conclude that leptogenesis
(at least in its simplest version,
where the flavour of the lepton $\alpha$ is summed over)
is not viable in our model.
Indeed,
this is a general result,
\textit{i.e.}\ it holds because of the texture zeros in our mass matrices
and not only in our specific model of the previous paragraph.
Since $M_R$ is diagonal,
we have $m_1 = |x|$ and $m_2 = |y|$.
We know that
\be
M_D = \sum_{b=1}^{n_H} Y_b v_b.
\ee
In order for the $M_D$ in equation~(\ref{md})
to follow from an \emph{Abelian} symmetry
which enforces its particular texture zeros,
the conditions
\begin{subequations}
\label{hgt}
\begin{eqnarray}
\left( Y_b \right)_{1\mu} &=& 0, \label{jboer} \\
\left( Y_b \right)_{2e} &=& 0, \label{lxpeq} \\
\label{vuycf}
\left( Y_b \right)_{1\tau} \left( Y_b \right)_{2\tau} &=& 0
\end{eqnarray}
\end{subequations}
[no sum over $b$ in equation~(\ref{vuycf})]
must hold for
all $b = 1, 2, \ldots, n_H$.
[If equation~(\ref{vuycf}) did not hold,
that would mean that $\nu_{R1}$ and $\nu_{R2}$
transform in the same way under the Abelian symmetry,
and then $\left( M_D \right)_{1\mu} = 0$
would imply $\left( M_D \right)_{2\mu} = 0$,
etc.]
Now,
equations~(\ref{hgt}) imply that
\be
\left( Y_b Y_b^\dagger \right)_{12}
= \sum_{\alpha = e, \mu, \tau}
\left( Y_b \right)_{1\alpha} \left( Y_b^\ast \right)_{2\alpha} = 0,
\ee
and therefore line~(\ref{line2}) vanishes.
Furthermore,
because of equations~(\ref{jboer}) and~(\ref{lxpeq}),
\be
\left( Y_b Y_c^\dagger \right)_{12}
\left( Y_c Y_b^\dagger \right)_{12}
=
\left( Y_b \right)_{1 \tau} \left( Y_c^\ast \right)_{2\tau}
\left( Y_c \right)_{1 \tau} \left( Y_b^\ast \right)_{2\tau},
\ee
which vanishes because of equation~(\ref{vuycf}).
Therefore,
line~(\ref{line1}) is zero and $\epsilon_{ib} = 0$.

\paragraph{Renormalization-group invariance:}
We now show that the conditions~(\ref{uviwe}) are,
in the context of the model that we have suggested in this Letter,
invariant under the (one-loop) renormalization-group (RG) evolution of $M$.
In that evolution,
one must consider~\cite{Grimus:2004yh},
in the context of a MHDM,
the effective operators
\be
\mathcal{O}_{bc} =
\sum_{\alpha, \beta = e, \mu, \tau}
\left( \phi_b^0 \nu_{L\alpha}^T - \phi_b^+ \alpha_L^T \right)
\kappa^{(bc)}_{\alpha \beta} C^{-1}
\left( \phi_c^0 \nu_{L\beta} - \phi_c^+ \beta_L \right)
\ee
where the $\kappa^{(bc)}$ are matrices in flavour space.
The effective operators $\mathcal{O}_{bc}$ arise in the seesaw mechanism
upon integrating out the heavy right-handed neutrinos.
One has $\kappa^{(bc)} = - Y_b^T M_R^{-1} Y_c$;
in our model,
from equations~(\ref{mr}) and~(\ref{wvuir}),
all the $\kappa^{(bc)}$ have,
at high energy,
vanishing $(e,\mu)$ and $(\mu,e)$ matrix elements.
During the RG evolution to lower energies,
the operators $\mathcal{O}_{bc}$ mix among themselves.
Moreover,
the RG derivative $\left. \mathrm{d} \kappa^{(bc)} \right/
\mathrm{d} \left( \ln{\mu} \right)$ contains~\cite{Grimus:2004yh}
matrices $\kappa^{(bc)}$
either multiplied on the right by matrices
\be
F_{bc} = \hat Y_b^\dagger \hat Y_c
\label{nvhuq}
\ee
or multiplied on the left by matrices $F_{bc}^T$.
In equation~(\ref{nvhuq}),
the Yukawa-coupling matrices $\hat Y_b$ are those which appear
in the Yukawa couplings of the right-handed charged leptons,
\be
\mathcal{L}_{\ell \, \mathrm{Yuk}} =
- \sum_{b=1}^{n_H} \sum_{\alpha, \beta = e, \mu, \tau}
\left( \hat Y_b \right)_{\alpha \beta}
\bar \alpha_R \left( \phi_b^- \nu_{L \beta} + {\phi_b^0}^\ast \beta_L \right)
+ \mathrm{H.c.}
\ee
In our model the matrices---\textit{cf.}\ equation~(\ref{jzirt})---
\be
\hat Y_1 = \mathrm{diag} \left( y_7, \, 0, \, 0 \right), \quad
\hat Y_2 = \mathrm{diag} \left( 0, \, 0, \, y_9, \right), \quad
\hat Y_3 = \mathrm{diag} \left( 0, \, y_8, \, 0 \right)
\ee
are diagonal and,
therefore,
the $F_{bc}$ are diagonal too.
Hence,
the multiplication of the $\kappa^{(bc)}$ by the $F_{bc}$
preserves the texture zeros that exist in the $\kappa^{(bc)}$.
Since in our model all the $\kappa^{(bc)}$
initially have null $(e,\mu)$ and $(\mu.e)$ matrix elements,
that feature will be preserved all along the RG evolution.
Therefore,
the condition~(\ref{m12}) is RG-stable.
As for the condition~(\ref{deter}),
it follows directly
from the fact that there are only two right-handed neutrinos,
and therefore it is stable independently of the specific form
of the RG equations.

\paragraph{Conclusions:}
In this Letter we have shown that the conditions~(\ref{uviwe})
on the effective light-neutrino Majorana mass matrix $M$
may be imposed by means of an Abelian symmetry $\mathbbm{Z}_8$
in a three-Higgs-doublet model with an extra scalar singlet.
We have furthermore demonstrated that those conditions agree
very well
with the known phenomenology
if the neutrino mass spectrum is inverted,
in which case
the Dirac phase $\delta$ is predicted to be extremely close to $\pm \pi / 2$.
The conditions~(\ref{uviwe}) do not work as well
if the neutrino mass spectrum is normal,
since in that case they necessitate a rather small $\theta_{13}$,
but they cannot yet be excluded
at the $3 \sigma$ level.
Unfortunately,
(flavour-independent) leptogenesis cannot work
in the scenario
in which the conditions~(\ref{uviwe})
are enforced by means of an Abelian horizontal symmetry.

\paragraph{Addendum:}
After completion of this work,
our attention has been called to a recent paper~\cite{Rodejohann:2012jz}
which has suggested,
albeit from a different starting point,
the conditions~(\ref{uviwe}).
That paper states that the condition~(\ref{m12})
``suffer[s] from little predictivity''
in the case of an inverted neutrino mass spectrum;
however,
that case is,
in our view,
the one in which the condition~(\ref{m12}) agrees best with experiment.

\paragraph{Acknowledgements:}
We thank Filipe Joaquim for a discussion on leptogenesis.
We thank both him and Werner Rodejohann for calling our attention
to some relevant literature.
The work of
P.M.F.\
is supported in part by the Portuguese
\textit{Fun\-da\-\c{c}\~{a}o para a Ci\^{e}ncia e a Tecnologia} (FCT)
under contract PTDC/FIS/117951/2010,
by the FP7 Reintegration Grant n.~PERG08-GA-2010-277025,
and by PEst-OE/FIS/UI0618/2011.
L.L.\ is funded by FCT through unit 777 and through
the projects CERN/FP/116328/2010,
PTDC/FIS/098188/2008,
and PTDC/FIS/117951/2010,
and also by the Marie Curie Initial Training Network ``UNILHC''
PITN-GA-2009-237920.

\end{document}